\documentclass[MSNbibl,nameyear,dvips]{arxstspdf}
\usepackage{flushend}
\usepackage{stfloats}
\volume{28}
\issue{1}
\pubyear{2013}
\firstpage{116}
\lastpage{130}
\doi{10.1214/12-STS405} 

\makeatletter

\newproclaim{definition}{Definition}

\newcommand{\rrVert}{\Vert}
\newcommand{\llVert}{\Vert}

\newcommand{\underset}[2]{\mathop{#2}_{#1}}

\newcommand{\mcF}{\mathcal{F}}
\newcommand{\mcI}{\mathcal{I}}
\newcommand{\mcL}{\mathcal{L}}
\newcommand{\mbR}{\mathbb{R}}
\newcommand{\mcX}{\mathcal{X}}
\newcommand{\bX}{\mathbf{X}}
\newcommand{\mcY}{\mathcal{Y}}
\newcommand{\bY}{\mathbf{Y}}

\newcommand{\bdelt}{\bolds{\delta}}
\newcommand{\beps}{\bolds{\varepsilon}}
\newcommand{\bbeta}{\bolds{\beta}}

\newcommand{\bone}{\mathbf{1}}

\newcommand{\E}{\mathrm{E}}
\newcommand{\Var}{\operatorname{Var}}
\newcommand{\argmin}{\arg\min}

\makeatother

\begin{document}
\begin{frontmatter}

\title{On Quantifying Dependence: A~Framework for Developing
Interpretable Measures}
\runtitle{On Quantifying Dependence}

\begin{aug}
\author[a]{\fnms{Matthew} \snm{Reimherr}\corref{}\ead[label=e1]{mreimherr@uchicago.edu}}
\and
\author[b]{\fnms{Dan L.} \snm{Nicolae}\ead[label=e2]{nicolae@galton.uchicago.edu}}
\runauthor{M. Reimherr and D. L. Nicolae}

\affiliation{University of Chicago}

\address[a]{Matthew Reimherr is Ph.D. Candidate,
Department of Statistics, University of Chicago,
5734 S. University Avenue, Chicago, Illinois 60637, USA \printead{e1}.}
\address[b]{Dan L. Nicolae is Professor, Departments of Statistics, Medicine and
Human Genetics, University of Chicago, Chicago, Illinois 60637, USA \printead{e2}.}

\end{aug}

%
\begin{abstract}
We present a framework for selecting and developing measures of
dependence when the goal is the quantification of a relationship
between two variables, not simply the establishment of its existence.
Much of the literature on dependence measures is focused, at least
implicitly, on detection or revolves around the inclusion/exclusion of
particular axioms and discussing which measures satisfy said axioms. In
contrast, we start with only a few nonrestrictive guidelines focused on
existence, range and interpretability, which provide a very open and
flexible framework. For quantification, the most crucial is the notion
of \textit{interpretability}, whose foundation can be found in the work
of Goodman and Kruskal [\textit{Measures of Association for Cross
Classifications} (1979) Springer], and whose importance can be seen in
the popularity of tools such as the $R^2$ in linear regression. While
Goodman and Kruskal focused on probabilistic interpretations for their
measures, we demonstrate how more general measures of information can
be used to achieve the same goal. To that end, we present a strategy
for building dependence measures that is designed to allow
practitioners to tailor measures to their needs. We demonstrate how
many well-known measures fit in with our framework and conclude the
paper by presenting two real data examples. Our first example explores
U.S. income and education where we demonstrate how this methodology can
help guide the selection and development of a dependence measure. Our
second example examines measures of dependence for functional data, and
illustrates them using data on geomagnetic storms.
\end{abstract}

%

\end{frontmatter}

\section{Introduction}
Exploring the relationships between variables is one of the most
fundamental tasks in statistics and at the heart of many statistical
analyses. A common goal is to clearly demonstrate the existence of
dependence between two variables.
Once the existence of a relationship is accepted or established,
dependence measures can be used to summarize that relationship in an
informative and concise fashion. They can provide deep insight into the
relationships between variables while being more easily communicated
than full model descriptions.
For example, in financial portfolios, the dependence between various
assets plays a crucial role in moderating risk. In epidemiology, it is
important to quantify
the dependence between diseases and various factors to determine which
are better predictors of risk. In genetics, the goal is often to
measure the strength of the
dependence between various phenotypes and genotypes to gauge which are
the most important biological pathways in the risk architecture of
complex traits. The dependence between genetic
markers plays a significant role in the design of association studies.
In any field where statistical procedures are applied, the
ability to quantify dependence in an interpretable fashion can be crucial.
Unfortunately, the proliferation of hypothesis testing has steered the
development of dependence measures away from interpretability. Many
modern measures are
developed with the goal of catching any trace of dependence in any
form, with less focus on the interpretability of their measures beyond
the extreme values of 0 and 1.
Examples of such measures that motivated our current work include the
distance correlation (\cite{szriba2007}; \cite{szri2009}), the maximal
information coefficient, MIC (\cite{reetal2011}) and copula based
measures (\cite{scwo1981}; \cite{sist2010}). Such measures are exciting
new tools for the detection of nonlinear relationships, but are
difficult to interpret at intermediate values.
The inability to interpret a measure of dependence is not necessarily
detrimental to an analysis, but it limits its use as a summary tool and
effectively isolates its utility to the realm of hypothesis testing or
the detection of dependence.

The goal of the present work is to develop a framework for dependence
measures when the primary task is the quantification and summarization
of a relationship
between two variables, not just the establishment of its existence. We
designed this framework with the aim of (a) helping practitioners
decide on an appropriate
dependence measure and consider more nonstandard measures if
applicable, (b) guiding the development of new measures of dependence
with interpretability as a
priority, and (c) starting a discussion challenging current views on
dependence. The central idea of the methodology is to build a
dependence measure by first constructing an appropriate measure of
information, determined by the practitioner and the setting, and then
using that measure of information to\vadjust{\goodbreak} quantify dependence.
Alternatively, for a preexisting measure, an interpretation can be
developed if one can find an information function embedded within it.
More succinctly, we adopt the view that measuring dependence in an
interpretable way is, in fact, about measuring the amount of relevant
information one variable contains about another.

To elucidate this dichotomy, and thus the need for our framework,
consider the high frequency data example on geomagnetic storms. We
will present this example with greater depth later on, but
understanding these storms has become very important
(see, e.g., \cite{mo2011}), as they can have damaging effects on GPS,
satellite, radar and data storage technologies. In that example we
measure the dependence between storms at different locations on the
earth with the goal of determining predictive capability and explained
variability. Since the storms are driven by solar wind, they are
obviously dependent, thus making a generic measure with no
interpretation of little use.

The notion of using information to measure dependence has been studied
extensively in the information theory literature and we reference
(\cite{as1990}; \cite{coth2006}; \cite{ebrahimi2010} and
\cite{gr2011}), to name only a few. Our perspective differs from the
information theory literature in two distinct ways. First, we do not
attempt to determine universally applicable or ideal information
functions and, in particular, we do not focus extensively on entropy,
though it will fall naturally within the proposed framework. Second,
we distinguish sharply between the detection of dependence and its
quantification. Only in the case of quantification do we insist on the
importance of an information function. More classical methods on
dependence measures go all the way back to \citet{re1959} where he
outlines a set of mathematical axioms that dependence measures should
satisfy. R\'{e}nyi's axioms have been modified in many various ways
(see, e.g., \cite{be1962}; \cite{ha1969}; \cite{scwo1981} and
\cite{ne2010}), but there is little discussion of one crucial
property: dependence measures intended for quantification should have
clear interpretations associated with them. Most of the common axioms
placed on dependence measures are less relevant in such a context.
Indeed, the only body of literature we could find developing similar
ideas was the seminal work of \citet{gokr1979}, where they meticulously
develop measures with probabilistic interpretations.\vadjust{\goodbreak} Our goal is
similar to theirs, but we achieve it in markedly different ways.

The paper is organized as follows. In Section~\ref{Fr} we outline our
framework for developing measures of dependence based on information
functions. In Section~\ref{Ex} we explore many examples of dependence
measures that fall nicely into the proposed framework. In
Section~\ref{Ap} we illustrate the importance of this framework in two
different real world settings. The first explores the relationship
between income and education and, in particular, explores how the
information relevant to the problem should guide the choice of measure.
The second application involves the analysis of geomagnetic storm data
and the relatively new area of functional data analysis. We show how
the ideas presented here can guide in the development of new
interpretable measures of dependence in that area. We conclude the
paper with a discussion in Section~\ref{disc}.

\section{Framework} \label{Fr}

An essential starting point in considering any dependence measure is
first examining its intended use. Though there may be many creative
applications for dependence measures, three of the most significant
ones are the following:
\begin{longlist}[(3)]
\item[(1)]\textit{detection}: detecting dependence in any form;
\item[(2)]\textit{ranking}: ordering the dependence in different
relationships (e.g., model selection);
\item[(3)]\textit{quantification}: summarizing a relationship in an
informative fashion.
\end{longlist}
A similar, though dichotomous, breakdown was noted by \citet{le1966}.

In the first setting, that a dependence even exists is sometimes
questionable. Thus, a measure leading to a valid and powerful testing
procedure would be most
desirable. As a simple illustration, suppose a researcher was examining
the relationship between income and height. In such a situation, there
is no clear reason, a priori, that the two variables should be
dependent. Thus, first using a measure designed for dependence
detection might be appropriate. Measures such as correlation can be
used to detect linear dependence, while the distance covariance or MIC
can be used for nonlinear relationships.

In the second setting, the main goal is to determine which
relationships are the strongest. For example, we may wish to rank or
select several variables that best explain income. Thus, we would aim
to choose a subset with the highest dependence.\vadjust{\goodbreak} In such a setting
statistical power and interpretability are not necessarily the primary
concern. The MIC, for example, attempts to establish a useful
``equitability'' property that assigns similar values to relationships
with similar noise levels, regardless of the functional nature of the
relationship.

In the third setting, the existence of a dependence is either obvious
or already well established. There, it would be more important to
quantify that dependence in a meaningful way. Using a similar simple
illustration, there is a clear, well-established dependence between
income and education. Thus, using a measure designed solely for
detection would be unproductive, while utilizing a carefully chosen
measure with a clear interpretation could provide a great deal of
insight that could be easily communicated to others.

We present our framework as a means of evaluating and developing
measures of dependence \textit{when the goal is the quantification of
a dependence}. We start by laying out guidelines or general properties
that dependence measures should satisfy in such a setting. We then
outline our method, based on incorporating information functions, to
demonstrate how to satisfy those guidelines.

\subsection*{Guidelines for Quantification}
In contrast to R\'{e}nyi, we propose only three guidelines instead of
six axioms:
\begin{longlist}[(3)]
\item[(1)]\textit{existence}: the measure should exist for a large
collection of random variables, vectors and/or functions, including
those relevant to the analysis;
\item[(2)]\textit{range}: the range of the measure should be $[0,1]$;
\item[(3)]\textit{interpretability}: the measure should have a clear
interpretation, for all possible values, based on information content.
Furthermore, $0$ should represent ``no information,'' while $1$
represents ``complete information.''
\end{longlist}
The difficulty in insisting on mathematical axioms is that
``interpretability'' is impossible to define mathematically, and yet is
the most crucial property. Our guidelines are designed to induce a
rather malleable framework that can easily adapt to the needs of the
researcher and the setting.

The first guideline simply indicates that the measure should be
applicable to any variables that one may come across in their analysis.
There is no reason why all measures should exist for all random
variables, or even all variables with some specific structure. The main
concern should be that the measure is at least well defined for the
possible variables that may arise in the analysis. The second guideline
simply creates a standard range of values for all measures. Since every
measure should have a concept of ``most dependent'' and ``least
dependent,'' it makes sense that the range be a bounded interval, and
using $[0,1]$ is fairly standard. Obviously some measures such as the
correlation can take negative values. However, only in the univariate
case where the relationships between variables are monotone does having
a signed measure make sense. In that case, nearly all signed measures
will have the same sign. So, for example, one could use correlation or
signed rank correlation to better understand the directional
relationship of the variables, while still using another $[0,1]$
measure to quantify the \textit{magnitude} of the dependence in a
relevant and interpretable way.

The essential guideline for quantification is ``interpretability.''
Nearly every dependence measure has a fairly clear interpretation at
its extreme values. In fact, a large emphasis is usually placed on the
interplay between complete independence/dependence\break and the extreme
values of the measure. However, we assert that not only should 0 and 1
have a clear interpretation, so should every value in between.
Furthermore, we claim that measuring dependence in an interpretable way
is really about measuring the amount of relevant information one
variable contains about another.

Noticeably absent are axioms/properties such as:
\begin{itemize}
\item zero dependence implies statistical independence,
\item symmetry,
\item invariance,
\item equivalence to absolute correlation in the joint normal setting.
\end{itemize}
The first property is important when detecting potentially nonlinear
relationships, but is not necessary for quantification, especially if
the interpretation of the measure is highly relevant.
By symmetry, we mean that the dependence between $X$ and $Y$ is
unchanged if the two are swapped. Models are often not symmetric, and
there is no reason to insist that all dependence measures should be.
However, we will discuss a potential method for symmetrizing in the
next section. Invariance means that one-to-one transformations of $X$
and/or $Y$ do not change the value of the measure. However, the scale
of $Y$ plays a crucial role in measures such as the correlation and
correlation ratio. Again, there is no obvious reason why all measures
should have such a property.

\subsection*{From Information to Dependence}
Incorporating the interpretability property is a challenging task. The
solution we adopt for building interpretability into a measure is based
on treating the
quantification problem as a user-specified information content exercise.
In particular, we introduce what we call an \textit{information link
function} that measures the amount of important information, as
determined by the practitioner
and the setting, one variable contains about another. Then a
practitioner could either build their own information link function or
select a predefined information
link function that emphasizes the priorities of their analysis. It is
important to note that even if a measure is based on an information
function, one still needs
to carefully examine the type of information to determine if it is
relevant. This is not a trivial task in applications, but should be a
main consideration in
evaluating a dependence measure. We will use $I(X,Y)$ to denote the
value of our information function evaluated at $X$ and $Y$, read as
``the
amount of information
$X$ contains about $Y$.''
%
\begin{definition}\label{linkfunc} Let $\mcF_2 \subset\mcF_1$ be
two collections of random variables, vectors and/or functions. We say
that a function $I$ is an \textit{information link function} over
$\mcF_1 \times\mcF_2$ if:
\begin{longlist}[(3)]
\item[(1)]$I\dvtx\mcF_1 \times\mcF_2 \to\mbR^+ $;
\item[(2)]$ I(X;Y) \leq I(Y;Y)$, for any $X \in\mcF_1$ and $Y\in\mcF_2$,
with $I(X;Y) = 0$ if they are independent;
\item[(3)] if, for any $X$ and $Z$ in $\mcF_1$, there exists a
function,~$f$, such that $Z = f(X)$, then $I(Z;Y) \leq I(X;Y) $ for
every $Y \in \mcF_2$.
\end{longlist}
\end{definition}
The first property simply indicates that information is a nonnegative
quantity. The second property indicates that a variable must contain
the maximum amount of information about itself and that independent
variables contain no information about each other. The third property
is a type of monotonicity and indicates that if one variable completely
determines another, then it must also contain more information.
A~consequence of the third property is that information link functions
are invariant under one-to-one transformations of the first argument
(assuming the transformation is in $\mcF_1$), but not the second. Such
a property is reasonable, as the scale of $Y$ can be crucial in
determining the scale of the measured information (such as explained
variance), however, one should not be able to obtain ``more
information'' about $Y$ by simply transforming the $X$ variable.

In contrast to the suggested guidelines, the information definition has
no interpretation requirement listed, as articulating such a property
mathematically is all but impossible. Instead, it is the responsibility
of the researcher to determine if a given information function has an
interpretation relevant to their analysis. Furthermore, those that
introduce new measures of dependence or information functions should
take care to explore and develop their possible interpretations.

Once a suitable information function is constructed, a dependence
measure is easily obtained via scaling. Define
\[
D(X,Y):= \frac{I(X,Y)}{I(Y,Y)},
\]
then $D$ satisfies the following:
\begin{longlist}[(5)]
\item[(1)]$D\dvtx\mcF_1 \times\mcF_2 \to[0,1] $;
\item[(2)]$ D(Y;Y) = 1$; if $X$ and $Y$ independent, then $D(X;Y)=0$;
\item[(3)] if, for any $X$ and $Z$ in $\mcF_1$, there exists a measurable
function, $f$, such that $Z = f(X)$, then $D(X;Y) \geq D(Z;Y)$ for every
$Y \in\mcF_2$;
\item[(4)] $D(X;Y)$ is invariant under one-to-one transformations of $X$
that stay in $\mcF_1$;
\item[(5)] built-in interpretability as a reduction or fraction of information.
\end{longlist}

Therefore, $D$ will satisfy all of our desired guidelines for
dependence measures and the task is reduced to determining an
appropriate information link function. Ideally, such a function will be
determined on a case-by-case basis as the practitioner and setting
dictate what information is of greatest importance.

Note that if symmetry of the measure is desired, there are at least
two potential methods of accomplishing it. However, for symmetry to be
coherent in our setting one would need to insist that $\mcF_1 = \mcF_2$
so that juxtaposing the variables makes sense. At that point, one
could symmetrize by either averaging the resulting $D(X;Y)$ and
$D(Y;X)$ or, more interestingly, by using an arithmetic mean
\[
D^S_1(X;Y) = \frac{I(X;Y) + I(Y;X)}{I(X;X) + I(Y;Y)}
\]
or a geometric mean
\[
D^{S}_2 (X;Y) = \sqrt{\frac{I(X;Y) \times I(Y;X)}{I(X;X) \times I(Y;Y)}}.
\]
In which case the measures could be interpreted as a kind of average
reduction in\vadjust{\goodbreak} information. The denominators above represent the ``total
information'' in the joint distribution.\vspace*{-2pt}

\section{Examples} \label{Ex}
We provide three examples that fall naturally into the proposed
framework. The first two, reflecting prediction and statistical
efficiency, are common in
statistics and actually constitute a large class of examples. The third
example, entropy, is more common in information theory, but
fits nicely into this framework as well.

\subsection*{Prediction}

One of the most common usages for exploring dependence is in the
prediction of or explaining the variability of a particular random
variable. For such a goal, we can start by building an information link
function that quantifies how knowing the value of one variable
increases the ability to predict another. As quantifying predictive
capability depends heavily on how one measures loss, we keep the
setting fairly general.

Let $g$ be a nonnegative penalty function, such that $g(0) = 0$; for
example, $g(x) = x^2$ would yield the usual $L^2$ prediction and $g(x)
= |x|$ the usual $L^1$ prediction. We start by defining an optimal
predictor of $Y$ based on~$X$. Since we will restrict our measure to
$\mcF_1 \times\mcF_2$, we only consider predictors of $Y$ that are
contained in $\mcF_1$. We assume that $\mcF_1$ at least contains all
of the constant values, that is, whatever space $Y$ is taking values in
is included in $\mcF_1$. So define, for $X$ taking values from $\mcX$
and $Y$ from $\mcY$,
\[
\hat{Y}(X) = \Bigl( \underset{f\dvtx\mcX\to\mcY, f(X) \in\mcF_1} {\arg
\min} E\bigl[g\bigl(Y - f(X)\bigr)\bigr] \Bigr) (X),
\]
that is, we choose a function of $X$ that best predicts $Y$, but also
falls into $\mcF_1$. See the \hyperref[app]{Appendix} for discussion on the existence
of such an estimate. We define $\hat{Y}_0$ to be the best constant
predictor of~$Y$. We can then quantify the increase in predictive
capability by examining the difference
\[
I(X;Y) = E\bigl[g(Y - \hat{Y}_0)\bigr] - E\bigl[g\bigl(Y -
\hat{Y}(X)\bigr)\bigr].
\]
The details showing that the above is a valid information link function
can be found in the \hyperref[app]{Appendix}. The resulting measure of dependence would
then be
\begin{eqnarray*}
D(X;Y) &=& \frac{I(X;Y)}{I(Y;Y)} \\
&=& \frac{E[g(Y - \hat{Y}_0)] - E[g(Y -
\hat{Y}(X))]}{E[g(Y - \hat{Y}_0)]},
\end{eqnarray*}
which can be interpreted as either the increase in predictive
capability, 0 being no increase and\vadjust{\goodbreak} 1 implying $Y$ is completely
determined by $X$, or as the proportion of ``g-variability'' of $Y$
explained by $X$.

For example, if $g(x) = x^2$, then the optimal predictor of $Y$ based
on $X$ is just $E[Y|X]$. The measure of dependence then becomes
\begin{eqnarray*}
D(X;Y) &=& \frac{E[(Y - E[Y])^2] - E[(Y - E[Y|X])^2]}{E[(Y - E[Y])^2]}\\
&=&
\frac{\Var(E[Y|X])}{\Var(Y)},
\end{eqnarray*}
which is just the well-known correlation ratio. Furthermore, if we
assume that the relationship between $X$ and $Y$ is linear, then
the above also equals the square of correlation.

\subsection*{Statistical Efficiency}
Another common statistical setting concerns what we call
\textit{statistical proxies}. Such objects arise when there is an ``optimal''
data set for inference on a particular parameter which is, for whatever
reason, not available and one has to use either the optimal set with
missing or coarsened values, or possibly a completely different data
set intended to be a substitute for the optimal one.
Thus, we call the observed data set a \textit{statistical proxy} for
the parameter of interest if the information it contains about the
parameter is redundant when the ``optimal'' set is also known. So
missing data and coarse data are special examples when the observed
data set contains only redundant information (as compared to the
complete data set) on the parameter of interest.
While observed variables are usually statistical proxies for the
complete ones, there are examples that are not based on missing data
from an optimal set. These include censored data, rounded measurements
and examples where variables of interest cannot be observed and
are replaced in inference by correlated variables. An example,
discussed in more detail at the end of this section, is that of genetic
association studies where causal variants are detected by testing
well-selected genetic markers.

So we wish to develop a measure of dependence when the goal is to
quantify how effective a statistical proxy one variable is for
another. The measures of dependence need to be tailored by the type of
statistical inference that is performed. We start by defining,
mathematically, what we mean by a statistical proxy for a parametric model.
%
\begin{definition}
We say that $X$ is a statistical proxy of $Y$ for a parameter $\theta$
if $Y$ is sufficient for $\theta$ with respect to the joint
distribution $\mcL(X,Y;\theta)$, that is, $\mcL(X| Y;\theta)$ is
almost surely constant with respect to $\theta$.
\end{definition}
This definition implies that $X$ contains only redundant information
for $\theta$ if $Y$ is known, though it gives no indication as to the
merits of $X$ as a proxy. For example, in a missing data problem, this
is obviously true as long as the missing data mechanism does not depend
on $\theta$, however, all that is required is that the missing data
contains only redundant information for $\theta$ when the full data
set is observed, which is a fairly mild assumption in most settings.

Performance of most likelihood based methods for estimation and
hypothesis testing can be evaluated, at least asymptotically, by the
Fisher information. For example, the asymptotic variance of the maximum
likelihood estimator is monotonically related to the Fisher
information, and so is the noncentrality parameter that drives the
power in the likelihood ratio test. So let $\mcF_2 = \{Y\}$ and $\mcF
_1$ be all proxies of $Y$. If the goal is to analyze the efficiency in
using $X$ for inference on~$\theta$, then a natural measure of
information is the Fisher information for $\theta$ based on~$X$:
\[
I(X;Y) = \mcI_X(\theta).
\]
The details showing that the above is a valid information link function
can be found in the \hyperref[app]{Appendix}. Since $X$ is a proxy for $Y$, it can
easily be shown that $\mcI_X(\theta) \leq\mcI_Y(\theta)$. Our
measure becomes
\[
D(X;Y) = \frac{ \mcI_X(\theta)}{ \mcI_Y(\theta)},
\]
which for estimation can be interpreted as the increase in variability
of the estimate or the decrease in accuracy when using $X$ in place of
$Y$. For hypothesis
testing it can be interpreted as the loss in power. In both cases, the
measure indicates the relative efficiency in inference about $\theta$
when using $X$
compared to $Y$. It is important to note that in the context of missing
data, the above measure is also closely related to the rate of
convergence for the EM
algorithm (see \cite{delaru1977}, for details).

As a very simple example, consider the case where $Y$ is normal with
mean $\mu$ and variance $\sigma^2$, and we are interested in
estimating $\mu$. However, suppose we only observe $X$ which is $Y$
with probability $p$ and missing (coded as 0) with probability $1-p$.
Put another way, $X = Y Z$, where $Z$ is Bernoulli with parameter $p$
and is independent of $Y$. Such a setting is usually called missing
completely at random or MCAR,
and the missing data mechanism is free of the parameter of interest
$\mu$. While we could technically compute the correlation, coding a
missing value as $0$ was completely arbitrary, thus, our measure should
not depend on that choice. As can be found in most graduate level
statistics text books, the Fisher information for $\mu$ with respect
to $Y$ is $1/\sigma^2$. The Fisher information for $\mu$ with respect
to $X$ is given by $p/\sigma^2$ (see the \hyperref[app]{Appendix} for details). Thus, in
this special case, the dependence measure becomes
\[
D(X;Y) = p.
\]
This is the well-known expected proportion of observed values, but it
also represents the relative efficiency in estimating $\mu$ when using
the $X$ observations in place of $Y$ observations.

There are situations where the main interest is in hypothesis testing,
and useful information link functions need to reflect easily
interpretable metrics such as the sample size necessary for achieving
some given type 1 and type 2 error rates. For example, in genome-wide
association studies, we only have data for single nucleotide
polymorphisms (SNPs) available on a particular genotyping array. Thus,
if a SNP is causal for a particular disease, but not on the array, its
signal could potentially be missed. However, if there is an arrayed
SNP highly correlated (called in \textit{linkage disequilibrium or LD})
with the causal SNP, then it can be used as a proxy for the causal
variant. The design of a genetic association study takes advantage of
the dependence between SNPs and of knowledge on how this dependence
affects the power of detecting associations. This knowledge is
quantified in a measure of dependence/LD, denoted by $r^2$, that has a
clear interpretation: it is approximately equal to the ratio of sample
sizes that leads to the same power when using the causal versus the
genotyped SNP. For example, suppose we would like to identify, in a
candidate gene study, if there exists a causal variant with a given
effect size. We can easily perform a power calculation for a causal SNP
that can specify the sample size, $n_1$, needed for detecting it.
Available are $n_2$ samples (with $n_2\geq n_1$), and the
interpretability of $r^2$ can help us in selecting an optimal
genotyping design: choose the minimum set of SNPs such that, for all
SNPs in the gene, there exist one in this set with pairwise $r^2\geq
n_1/n_2$. Note that measures based on the idea of asymptotic relative
efficiency (ARE) are not the only way one could design interpretable
functions in hypothesis testing. For example,\vadjust{\goodbreak} one can use elements in
the distribution of the likelihood ratio statistic to quantify impact
on power (see \cite{nimeko2008}; \cite{reni2011}).

The idea of sample size as a measure of information for exchangeable
(e.g., i.i.d.) data could be a powerful tool for translating attributes of
joint distributions
to applied scientists. There are many situations where the interest is
in observable claims (length of confidence intervals, precision of
estimation, power of a statistical test, etc.) on the marginal
distribution of $Y$, that is, in objects that can be calculated from
the distribution function of $Y$. For example, we could be interested
in a quantile of $Y$ (the percentage of households in a city with
annual income larger than \$250K) and we would like to express that
with a narrow confidence interval (length smaller than one percent).
Using information on the distribution of $Y$ (e.g., national data for
income), we can predict the sample size necessary for the needed claim,
the issue being that we could collect data only for a proxy, $X$ (such
as the tax rate for the household). Obviously, we need a larger sample
size to obtain the same width for a confidence interval, and the ratio
of these two sample size offers an easily interpretable measure of dependence.

\subsection*{Entropy}
Entropy is widely used in the information theory literature as the
primary measure of information content in a random variable
(see, e.g., \cite{coth2006}) and arises in the statistics literature as
the expected value of the log likelihood. The interpretability of the
entropy is a bit questionable except for in some very specific
circumstances, but it is nevertheless very popular in the field of
information theory. Thus, it may be reasonable that a practitioner
would choose entropy as the measure of information they are concerned
with, and, in particular, how knowing the value of one variable reduces
the entropy in another.

The entropy of a random variable $Y$---for simplicity, we assume that
$Y$ is discrete---with probability mass function $f_Y$ is defined as
\[
H(Y) = - \E\bigl[\log\bigl(f_Y(Y)\bigr)\bigr].
\]
And the conditional entropy of $Y$ given $X$ is
\[
H(Y|X) = - \E\bigl[\E\bigl[\log\bigl(f_{Y|X}(Y|X)\bigr)|X\bigr]\bigr],
\]
which indicates the entropy of $Y$ given $X$, averaged over $X$. Thus,
using the reduction in entropy of $Y$ by knowing $X$ as our measure of
information gives
\[
I(X;Y) = H(Y) - H(Y|X),
\]
which is commonly called the \textit{mutual information} in $Y$ about
$X$. The details showing that the above is a valid information link
function can be found in \citet{coth2006}. This yields the dependence
measure
\[
D(X;Y) = \frac{H(Y) - H(Y|X)}{H(Y)},
\]
which is easily interpreted as the proportional reduction in entropy of
$Y$ by knowing $X$ (see also \cite{ebrahimi2010}). It is maybe
interesting to note that while the mutual information is symmetric, the
above dependence measure is not.

\section{Applications} \label{Ap}
In this section we discuss applications to income and education, and
geomagnetic storms. While these examples are more focused on
prediction, additional examples involving statistical efficiency and
missing information in genetic association studies can be found in
\citet{ni2006}, \citet{nimeko2008} and \citet{reni2011}.

\subsection*{Income and Education}
One of the driving motivators of pursing education is the potential for
higher income. A vast amount of data and reports exploring
their relationship can be found on the website for the Department of
Labor Statistics \href{http://www.bls.gov}{www.bls.gov}. In this example we will explore the
differences between men and women in terms of how their incomes are
affected by education levels. Furthermore, we will demonstrate how the
choice of dependence measure can play a crucial role in understanding
and communicating that difference. The data we explore here consists of
approximately 1.25 million individuals living in the U.S. aged 25 and
over and receiving an annual income (see \texttt{%
\href{http://factfinder.census.gov/home/en/acs\_pums\_2009\_1yr.html}{http://factfinder.census.gov/home/en/}\break
\href{http://factfinder.census.gov/home/en/acs\_pums\_2009\_1yr.html}{acs\_pums\_2009\_1yr.html}}
for
further details).

How to choose a dependence measure for such a setting is not completely
obvious. Classical choices include correlation (assuming education is measured
quantitatively) and the correlation ratio in conjunction with a
generalized linear model. However, individuals often strive to hit
certain income thresholds. At
low levels of income, individuals may try to make it out of poverty or
above minimum wage. Thus, a very meaningful question would be, how does
education affect the
chances of making it past a certain income threshold? For this example,
we will use a threshold of \$35,000,\vadjust{\goodbreak} the approximate median income of
U.S. adults over 25
years of age (see \cite{us2010}).

Let $Y$ be an indicator variable equaling $1$ if an individual's
personal income is over \$35,000 and 0 otherwise. Let $X$ be the
education level of that individual, equaling $0,1,2$ and $3$
representing education levels of ``less than high school,'' ``high
school degree or equivalent,'' ``bachelor's or associates degree'' and
``higher degree,'' respectively. See \citet{ef1978} for a
discussion on dependence measures for binary response variables.
While the literature on binary data is quite large, we also cite
\citet{gokr1979}, \citet{mcne1989}, \citet
{lilaha1991} and
\citet{liqaze1992} and the references therein. The last two references
deal especially with odds ratios which we have not touched on here.
Here we will compare 3 different measures of dependence: the
correlation ratio (sometimes called Efron's $R^2$ in this setting), the
ratio of reduction in deviance and the ratio of reduction on 0--1
prediction error. The first measure might be considered the most
natural generalization of the $R^2$ from linear regression and gives
the reduction in $L^2$ prediction error. The second measure is
commonly used in the theory of generalized linear models (see
\cite{mcne1989}). The third measure is natural because $Y$ takes
discrete values, so we can ask what the probability is that we
incorrectly predict $Y$. As an aside, the $D_3$ measure can also be
viewed as the reduction in $L^1$ prediction error, as that measure will
coincide with $D_3$ in this case. Let $Y_1, \ldots, Y_n$ be the binary
incomes for the $n$ individuals in the data set, and let $X_1, \ldots,
X_n$ be their corresponding education levels.\vspace*{1pt} Define two fitted values
as $\hat{Y}_i = \hat{E}[Y_i | X_i]$ and $\tilde{Y}_i = I\{\hat{Y}_i
\geq0.5\}$. Also define $\hat{Y} = n^{-1} \sum Y_i$ and $\tilde{Y} =
I\{\hat{Y} \geq0.5\}$ which correspond to the unconditioned fitted
values. Then the measures of dependence can be expressed as
\begin{eqnarray*}
&&
\mbox{correlation ratio:} \\[-2pt]
&&\quad\hat{D}_1(X,Y) = 1 - \frac{\sum(Y_i -
\hat{Y}_i)^2}{\sum(Y_i- \hat{Y})^2},
\\[-2pt]
&&
\mbox{deviance ratio:} \\[-2pt]
&&\quad\hat{D}_2(X,Y) \\[-2pt]
&&\qquad= 1 - \sum\bigl[
Y_i \log(Y_i/ \hat{Y}_i) \\[-2pt]
&&\hspace*{69pt}{}+ (1-Y_i) \log\bigl((1-Y_i)/(1-\hat{Y}_i)\bigr)
\bigr]\\[-2pt]
&&\hspace*{19pt}\quad\qquad{}\big/\sum\bigl[ Y_i \log(Y_i/ \hat{Y})\\[-2pt]
&&\hspace*{77pt}{} + (1-Y_i) \log
\bigl((1-Y_i)/(1-\hat{Y})\bigr) \bigr],
\\[-2pt]
&&
\mbox{0--1 ratio:}\\[-2pt]
&&\quad \hat{D}_3(X,Y) = 1 - \frac{\sum I\{Y_i \neq
\tilde{Y}_i\}}{\sum I\{Y_i \neq\tilde{Y}\}}.\vadjust{\goodbreak}
\end{eqnarray*}
Notice that in this context, the $D_3$ measure has a very useful and
relevant interpretation. Since we are looking at an income threshold of
\$35,000, we are especially interested in how well education
predicts being over or under that threshold. $D_3$ gives a literal
measure of reduction in prediction error that most people can
understand: if $D_3 = 0.30$, then knowing a person's education level
decreases the chances of incorrectly predicting them above/below
\$35,000 by $30\%$ (as compared to using the population average). The
measure $D_2$, on the other hand, is very difficult to interpret. It
gives a reduction in a log type penalty, but it is difficult to give it
much more of an interpretation (although for statisticians they can
view it as a reduction in the expected log likelihood). The measure
$D_1$ has a bit more of an interpretation and closely resembles the
regression $R^2$, but considering the discrete nature of $Y$, it is
difficult to explain why one should be especially interested in an
$L^2$ type loss.

The fitted values are computed using the full model and we compare the
different measures in males and females. The results are summarized in
Table~\ref{tinc}. Note that even the smallest gender/income/education
group has over 80,000 individuals, making all model estimation error
essentially negligible. As we can see from the table, each measure
differs across gender, however, the magnitudes of the measures are
quite different. While $D_1$ and $D_2$ are approximately $40\%$ higher
for females than males, $D_3$ is almost $100\%$ higher.

\begin{table}
\caption{Income and education measures of dependence. $D_1$, $D_2$ and
$D_3$ correspond to the correlation ratio, the defiance ratio and the
0--1 loss ratio, respectively} \label{tinc}
\begin{tabular*}{\tablewidth}{@{\extracolsep{\fill}}lccc@{}}
\hline
\textbf{Gender} & $\bolds{D_1}$ & $\bolds{D_2}$ & $\bolds{D_3}$ \\
\hline
Male & $0.1092$ & $0.0843$ & $0.1171$ \\
Female & $0.1522$ & $0.1183$ & $0.2329$ \\
\hline
\end{tabular*}
\end{table}

To further understand these relationships, consider the conditional
probabilities $P(Y = 1 |X)$ for different levels of education as given
in Table~\ref{tinc2}. The difference between men and women is fairly
remarkable, especially at lower education levels. For those with less
than a high school degree ($X=0$), men are more than twice as likely
than women to make more than the median income. This trend levels off
at higher education levels, as among those with a graduate level degree
$(X=3)$, men are only $12\%$ more likely than women to make more than
the median income.

\begin{table}
\caption{The probabilities of making more than the median income~given
the education level $X$ for males and females.\break Note $p(y|x) = P(Y=y|
X=x)$} \label{tinc2}
\begin{tabular*}{\tablewidth}{@{\extracolsep{\fill}}lccccc@{}}
\hline
\textbf{Gender} & $\bolds{p( 1 | 0)}$ & $\bolds{p( 1 | 2)}$
& $\bolds{p( 1 | 2)}$ & $\bolds{p( 1 | 3)}$ &
$\bolds{P(Y=1)}$ \\
\hline
Male & 0.2719 & 0.5310 & 0.7404 & 0.8351 & $ 0.6046 $\\
Female & 0.0836 & 0.2761 & 0.5462 & 0.7445 & $0.4140 $ \\
\hline
\end{tabular*}
\end{table}

The differences between men and women in regards to the
income/education relationship is remarkable. The measure of dependence
$D_3$ picks up this difference more clearly than $D_1$ and $D_2$ and
has a very relevant interpretation as the reduction in literal
prediction error. After seeing the conditional probabilities in
Table~\ref{tinc2}, it is easy to understand why education plays a
larger role for women than in men; by attaining higher levels of
education, women are able to significantly lower this income gap
compared to men.

\subsection*{Geomagnetic Storms}
Here we present an example of how our methodology can help guide the
development of new measures of dependence. The magnetosphere of the
earth forms part of the exosphere, the earth's atmosphere's outermost
layer. Solar wind emitted by the Sun is directed around the earth by
the magnetosphere, but the interaction of the two generates a
tremendous amount of electrical current and electromagnetic activity.
Solar flares can generate strong geomagnetic substorms, an example of
which is the Aurora Borealis. Particles from the solar wind make it to
the innermost layer of the magnetosphere, called the Ionosphere, and
ionize the gases, causing an amazing display of light. Substorms
typically last one or two days and can be very disruptive to global
positioning systems and radio and radar technologies that bounce their
signals off the the ionosphere, as well as outright damaging
satellites, power grids and data storage technologies. This topic has
gained a great deal of attention recently, as we are approaching the
peak of the solar magnetic activity cycle.

Understanding the nature of these storms is an important goal, but one
made difficult by the fact that the magnetosphere is too low for
satellites, but too high for aircraft. To that end, INTERMAGNET is a
network of terrestrial observatories that monitor eletromagnetic
activity in the magnetosphere and attempt to provide almost real time
data on the geomagnetic activity at their location. A large scale
analysis of their data is far beyond the scope of this paper. Instead,
we focus on measurements taken in College (coded as CMO), Alaska and
Honolulu (coded as HON), Hawaii in 2001. The data consists of 120 days
where storms occurred, taken from January through September. We further
separate the data into three sets of 40 pairs. The first set consists
of storms in January through March, the second set April through June,
and the third set July through September. On each day, 1440 equally
spaced measurements are given, making more traditional methods
difficult to apply. Each value is measured in nanoteslas and only
indicates the strength of the horizontal component of the magnetic
field. Time is measured in terms of Universal Time (UT). We view each
day as a single functional observation because of the daily rotation of
the earth. By ``functional observation'' we mean that we treat the curve
from a particular day as an observation from a random function taking
values in a function space. Such an approach is commonly called
\textit{functional data analysis} or FDA for short.

A long term goal would be the ability to predict future storm activity
at different locations on the globe, using data from other stations.
For example, since the substorms are driven by the sun, a station's
storm activity usually dies out a night. Thus, we may potentially use
stations currently facing the sun to predict the next days storm
activity for stations currently facing away from the sun. So we build
measures based on $L^2$ prediction, that is, how well we can predict
the storm activity in one station (HON) given that we observe the
activity at another (CMO). A larger analysis would use data from
multiple stations as well as taking care with the differing time zones
(CMO is only two hours ahead of HON, so one really cannot use an entire
CMO day to predict a HON day), but our approach will be sufficient
enough to illustrate our dependence framework.

We assume that $Y$ and $X$ are random functions taking values from
$L^2[0,1]$, representing the entire curve of values measured in HON and
CMO, respectively, on a particular day. As was said, the information we
are concerned with is $L^2$ prediction. Thus, we can take the reduction
in $L^2$ prediction as our measure of information
\[
I(X;Y) = E\bigl\| Y - E[Y] \bigr\|^2 - E\bigl\| Y - E[Y|X] \bigr\|^2.
\]
We should note that above $Y, X, E[Y]$ and $E[Y|X]$ are all functions
and $\| \cdot\|$ is the functional $L^2$ norm (the domain depends on
how you parametrize time over a day, but is usually taken to be $[0,1]$
for simplicity). And we arrive at a kind of functional version of the
correlation ratio
\[
D_1(X;Y) = \frac{ E\| Y - E[Y] \|^2 - E\| Y - E[Y|X] \|^2}{E\| Y -
E[Y] \|^2}.
\]
As in the univariate case, $D_1$ can be interpreted as explained
variability or reduction in $L^2$
prediction error. This measure is given (\cite{rasi2005}) in the
context of a goodness-of-fit measure for the functional linear model
(which we will use to estimate the measures here).

While $D_1$ has a very nice interpretation, one possible concern could
be that $D_1$ will be heavily influenced by coordinates of $Y$ with
high variability. This can be viewed positively in certain situations,
as more variability means that there is more to explain. But, as with
our example here, the variability will naturally change depending on
the time of day, as the magnetic activity is driven by the sun. Thus,
we may want a measure that takes the changing variability into account
and is not as influenced by the more variable coordinates. With that in
mind, we propose two additional measures of dependence:
\begin{eqnarray*}
&&
D_2(X;Y) \\[-2pt]
&&\quad = \frac{ E\llVert ({Y - E[Y]})/{S} \rrVert^2 - E\llVert
({Y - E[Y|X]})/{S} \rrVert^2}{E
\llVert ({Y - E[Y]})/{S} \rrVert^2} \\[-2pt]
&&\quad= 1 - E\biggl\llVert
\frac{Y - E[Y|X]}{S} \biggr\rrVert^2,
\\[-2pt]
&&
D_3(X;Y) \\[-2pt]
&&\quad = 1-\frac{E [(\bY- E[\bY| X])^T \Sigma^{-1}(\bY-
E[\bY| X]) ]}{ E [(\bY- E[\bY])^T \Sigma^{-1}(\bY-
E[\bY] ) ] }
\\[-2pt]
&&\quad = 1 - \frac{E [(\bY- E[\bY| X])^T \Sigma^{-1}(\bY- E[\bY|
X]) ]}{d},
\end{eqnarray*}
where $S(t) = (\Var(Y(t))^{1/2}$, $\bY\in\mbR^d$ is the projection of
$Y$ onto the $d$ most significant principal components, and $\Sigma$ is
the variance--covariance matrix of $\bY$. Here, $D_2$ can be
interpreted as the explained variability averaged over time and it
simplifies in the above way because the time interval is $[0,1]$
(otherwise it would be scaled by the length of the interval). Notice
that this is an average of the coordinate-wise measure given in
\citet{rasi2005}. The third measure, $D_3$, which we have not
seen in
previous FDA literature, utilizes a principal component analysis to
project the data onto a finite dimensional setting,\vadjust{\goodbreak} denoted $\bY$, and
then computes a multivariate goodness-of-fit measure there. Again, the
effect now is that the measure gives an average goodness of fit, but
this time averaged over the principal components. This interpretation
becomes especially clear when one takes into account that $\Sigma$ is
in fact a diagonal matrix. {Commonly, one chooses the number of
principal components so that a large percentage of the variability,
say, 85--95\%, is explained by the PCs. When choosing the number of
PCs for $X$ one could also use a cross-validation in terms of
predicting $Y$. In our examples we always choose the number of PCs
such that 85\% of the variability is explained.} Both $D_2$ and $D_3$
attempt to ``average out'' larger components that might otherwise
dominate the measures, but they do it in very different ways. The
measure $D_2$ averages the dependence over time, thus smoothing out
more variable time periods, while $D_3$ averages over the components so
that larger components do not completely dominate the measure. The
appropriateness of the measures will depend on the setting, though in
most cases $D_1$ will be very natural.

\begin{table}
\caption{The FDA measures of dependence for magnetic
storm data evaluated over different seasons. Here $X$ represents
Honolulu, Hawaii and $Y$ represents College,
Alaska}\label{tb1}
\begin{tabular*}{\tablewidth}{@{\extracolsep{\fill}}lccc@{}}
\hline
\textbf{Period} & \textbf{Jan.--March} & \textbf{April--June} & \textbf{July--Sept.} \\
\hline
$\hat{D}_1(Y, X)$ & 0.748 & 0.576 & 0.594 \\
$\hat{D}_2(Y, X)$ & 0.597 & 0.569 & 0.453 \\
$\hat{D}_3(Y, X)$ & 0.511 & 0.582 & 0.427 \\
\hline
\end{tabular*}
\end{table}

Table~\ref{tb1} gives estimates of the dependence measures over the
three different seasons. The story\break changes slightly depending on the
measure, which makes the interpretations all the more relevant. The
first measure is strongest in January through March and weaker in the
other two seasons. The second measure decreases with each season, with
a much larger drop moving from the second to third season. The third
measure indicates that the dependence is actually strongest in April
through July, though agrees with $D_2$ in that the July through
September dependence is weakest. Since $D_3$ averages over principal
components, the implication would be that the fit for the first
component is much better in the January through March storms (since
$D_1$ is so much larger for that season) as compared to the other two
seasons, while the second season has a better fit for the later components.

Our findings go a step beyond ranking of the\break strength of the dependence
across seasons.\vadjust{\goodbreak} Since our dependence measures were built upon a measure
of information, each number in Table~\ref{tb1} also has deeper meaning
beyond ordering. Here the first measure can be interpreted as the
percentage of $L^2$ variability in one station explained by observing
another. For example, in January through March, almost $75\%$ of the
variability in a Hawaiian storm is accounted for by observing the
corresponding storm in Alaska. What this means for scientists is that,
after taking into account a storm in Alaska, only $25\%$ of the energy
in the Hawaiian storm is still unpredictable or unaccounted for. The
second measure gives the average ``over time variability'' of one
station explained by another, while the third measure gives the average
``over principal components'' variability explained.


\section{Discussion} \label{disc}

In this paper we present a framework for developing and analyzing
measures of dependence when the goal is to explore and summarize the
relationship between two
variables. The framework consists of just a few guidelines and an
information-based methodology designed to achieve those guidelines. We
demonstrated how many well-known
measures fit into this framework and presented two real data examples
that demonstrated two distinct ways in which this methodology could be
used. The first
example was based on income and education and showed how the context of
the problem and the goals of the researcher should dictate the chosen
dependence measure.
The second example developed a new measure of dependence for functional
data. The measure was developed to ensure that it was not only
interpretable and informative, but that the information it conveyed was
highly relevant to the context of the problem.

The present work is in the same vein as the work of Goodman and
Kruskal, but we go about it in markedly different ways. While they
focus on measures with probabilistic interpretations, we exploit more
general measures of information. In both cases, though, the goal is to
develop measures with useful and relevant interpretations. A
significant motivator for the present work was the development of tools
such as the distance correlation (see, e.g.,
\cite{szriba2007}; \cite{szri2009}), the maximal information coefficient
(\cite{reetal2011}) and copula-based measures (see, e.g.,
Schwei\-zer and Wolff, (\citeyear{scwo1981}); \cite{sist2010}). Such measures provide interesting and
powerful methods for\vadjust{\goodbreak} detecting nonlinear dependence, but are very
difficult to interpret. For example, saying that the correlation
between two variables is $0.5$ can have numerous useful
interpretations, some of which are classical (explained linear
variability) and some of which are less standard (Reimherr and Nicolae\break (\citeyear{reni2011})).
However, knowing that the distance correlation or the MIC between two
variables is 0.5 currently gives us relatively little insight into the
relationship between two variables.

The link between dependence measures and types of information will
hopefully help open up an array of different types of dependence
measures for any given setting.
While researchers can always choose more classical measures based on
prediction or entropy, it is important for them to know that
alternative measures with
distinctly different meanings are also available or can be constructed.
For example, in the case of missing information, dependence measures
related to the
fraction of missing information can be constructed along the lines of
our simple example. In the case of hypothesis testing, measures related
to the relative
efficiency of tests based on two different variables can be
constructed. Such a measure can be interpreted as a ratio of sample
sizes that yield the same
statistical power, a very attractive interpretation. Hopefully, more
work will follow that shows how other important quantities can be used
to construct new,
nonstandard, yet very informative measures of dependence.

It is worth noting that the present work focuses almost exclusively on
developing theoretical measures based on joint distributions. The
issue of estimation is left almost completely untouched, as that in and
of itself is a fairly complex problem. Moment estimators can easily be
used to estimate measures such as correlation, however, traditional
estimation of the correlation ratio or entropy-based measures require
an assumption about the joint distribution of the two variables, which
can be a nontrivial problem. Nonparametric estimation of the
correlation ratio can be found in \citet{dosa1995}, however, it is
unclear whether more general nonparametric estimates of more general
information functions can be developed, or if estimation must be done
on a case-by-case basis. {Clearly, one should take great care when
attempting to apply a measure whose estimation and inferential
properties are not well established.}

{We have also not discussed the important concept of conditional
dependence measures, which would be useful, for example, when one has a
fair amount of collinearity between explanatory variables. We believe
one could adjust the current framework to handle conditional dependence
measures by adjusting the spaces $\mcF_1$ and $\mcF_2$ to be
conditional random variables. Of course, practically one needs a
measure where one can easily take a conditional expectation or be able
to work with conditional distributions. Such a step can be accomplished
with nearly all measures presented here. However, given the importance
of the problem, we refrain from exploring the issue further presently.}

Our hope is that the current work will start a discussion on measures
of dependence. Whether it be in new research or in the classroom, we
believe the interpretation of a measure should always be emphasized. It
allows researchers to better determine the relevance of a measure to
their analysis, giving clear interpretations to help cultivate their
conclusions, as well as providing intuition and understanding to
students and nonstatisticians.

\begin{appendix}\label{app}
\section*{Appendix}
\subsection*{Prediction}
Here we show that the information link functions given in the
prediction section satisfy the assumptions in Definition~\ref{linkfunc}.
Assume that $X$ takes values from a set $\mcX$ and $Y$ from $\mcY$.
Furthermore, assume both sets are separable Banach spaces. Let $\mu_X$,
$\mu_Y$ and $\mu_{X,Y}$ be the probability measures induced by
$X$, $Y$ and $(X,Y)$, respectively. By definition, property 3 for
information link functions is satisfied. Since all constants are
included in $\mcF_1$, $I(X;Y)$ is positive and property 1 is
satisfied. Since $Y$ predicts itself perfectly and we assume $g(0) =
0$, then the first part of property 2 is satisfied. To see\vspace*{-0.5pt} that
$I(X;Y)$ is zero when $X$ and $Y$ are independent, we start\vspace*{2pt} by showing
that any predictor based on $X$ cannot do better that $\hat{Y}_0$.
Consider,\vspace*{-0.5pt} for any $f$ such that $f(X) \in\mcF_1$,
\[
E\bigl[g\bigl(Y - f(X)\bigr)\bigr] = \int_{\mcX\times\mcY} g\bigl(y -
f(x)\bigr) \,d \mu(x,y).
\]
Since $g$ is nonnegative (we of course have to assume $g$ is
measurable as well), the integral exists and by Fubini's theorem, when
$X$ and $Y$ are independent, equals
\begin{eqnarray*}
&&
E\bigl[g\bigl(Y - f(X)\bigr)\bigr] \\
&&\quad= \int_{\mcX} \biggl(\int
_{\mcY} g\bigl(y - f(x)\bigr) \,d \mu_Y(y) \biggr)
\,d \mu_X(x) \\
&&\quad= \int_{\mcX}\E\bigl[g\bigl(Y - f(x)
\bigr)\bigr] \,d \mu_X(x).
\end{eqnarray*}
Since $\hat Y_0$ is the best constant predictor of $Y$, we have that
$\E[g(Y - f(x))] \geq\E[g(Y - \hat{Y}_0)]$, for all $x \in\mcX$
and, therefore,
\begin{eqnarray*}
E\bigl[g\bigl(Y - f(X)\bigr)\bigr] &\geq&\int_{\mcX}\E\bigl[g(Y
- Y_0)\bigr] \,d \mu_X(x) \\
&=& \E\bigl[g(Y -
Y_0)\bigr].
\end{eqnarray*}
Thus, any predictor based on $X$ cannot do better than $\hat{Y}_0$ and
we obtain $0 \leq I(X;Y) \leq I(Y_0; Y) = 0$ if $X$ and $Y$ are
independent and, therefore, all 3 properties are satisfied.

Note that the existence of the estimators $\hat{Y}$ and $\hat{Y}_0$ can
be guaranteed by placing some requirements on $g$ and $\mcF_1$. If $g$
is a continuous convex function and $\mcF_1$ is a finite dimensional,
closed and convex set, then the existence of a solution follows from
standard convexity theory. If $\mcF_1$ is infinite dimensional, then
one also needs that $g$ is coercive to guarantee that a solution
exists. For more details see any text on convex optimization or
variational calculus (e.g., \cite{gefo1963}; \cite{boli2004}).

\subsection*{Relative Efficiency}
Here we show that the information link functions given in the relative
efficiency section satisfy the assumptions in Definition \ref
{linkfunc} and detail the calculations for the MCAR example. Assume
that the joint and marginal distributions are continuously
differentiable. Property 1 for information link functions is satisfied
by definition. To establish property 2 consider that
\[
\mcI_{X,Y}(\theta) = \mcI_{Y|X}(\theta) + \mcI_{X}(
\theta) = \mcI_{X|Y}(\theta) + \mcI_{Y}(\theta).
\]
Since $Y$ is sufficient for $\theta$, $f(X|Y;\theta)$ is constant
with respect to $\theta$ and we have
\[
\mcI_{X|Y}(\theta) = \E_\theta\biggl[ \biggl(
\frac{\partial
}{\partial\theta}\log f(X|Y;\theta) \biggr)^2 \biggr] = 0.
\]
Thus, $\mcI_{X}(\theta) \leq\mcI_{Y}(\theta)$, which proves
$I(X;Y) \leq \break I(Y; Y)$. If $X$ and $Y$ are independent, then
$f(X|Y;\break\theta) = f(X;\theta)$ and is constant with respect to $\theta
$. Thus, $\mcI_X(\theta) = 0$ and the second property is established.
The third property now follows from property 2 since $Z$ is a proxy of
$X$ for $\theta$.

For the MCAR example, the Fisher information for $\mu$ with respect to
$X$ can be computed as
\begin{eqnarray*}
\mcI_{X}(\mu) & = & - \E\biggl[ \frac{\partial^2}{\partial\mu^2} \log
\bigl(f
\bigl(X,Z; \mu, \sigma^2, p\bigr)\bigr) \biggr]
\\
& = & - \E\biggl[ \frac{\partial^2}{\partial\mu^2} \log\bigl(f\bigl(X,0;
\mu,
\sigma^2, p\bigr)\bigr) \Big| Z = 0 \biggr]\\
&&{}\cdot(1-p)
\\
&&{} - \E\biggl[ \frac{\partial^2}{\partial\mu^2} \log\bigl(f\bigl
(X,1; \mu,
\sigma^2, p\bigr)\bigr) \Big| Z = 1 \biggr]p.
\end{eqnarray*}
If $Z=0$, then $X=0$ with probability 1 and
\[
- \E\biggl[ \frac{\partial^2}{\partial\mu^2} \log\bigl(f\bigl(X,0; \mu
, \sigma^2,
p\bigr)\bigr) \Big| Z = 0 \biggr](1-p) = 0.
\]
If $Z = 1$, then $X$ is normal with mean $\mu$ and $\sigma^2$ which means
\[
- \E\biggl[ \frac{\partial^2}{\partial\mu^2} \log\bigl(f\bigl(X,0; \mu
, \sigma^2,
p\bigr)\bigr) \Big| Z = 1 \biggr]p = p/\sigma^2.
\]

\subsection*{Income and Education}
Here we detail how the 0--1 information link function from the income
and education example was derived, as well as proving that it satisfies
the assumptions in Definition~\ref{linkfunc}. We can take $\mcF_2$ to
be the set of Bernoulli random variables and $\mcF_1$ to be the set of
all random variables and vectors. Let $\hat{Y}$ be a predictor of $Y$
based on no information (not conditioned on any other random variables)
and $\hat{Y}(X)$ based on $X$. If we evaluate $\hat{Y}$ based on 0--1
loss, then our error is
\begin{eqnarray*}
\E[\bone_{\hat{Y} \neq Y}] &=& P(\hat{Y} \neq Y) = P(\hat{Y} \neq0| Y =
0)P(Y=0) \\
&&{}+
P(\hat{Y} \neq1| Y = 1)P(Y=1).
\end{eqnarray*}
Since $\hat{Y}$ is based on ``no information,'' it must be independent
of $Y$. Thus, if we let $p_y$ denote the $P(Y=1)$ and $q_y = P(\hat{Y}
= 1)$, then
\begin{eqnarray*}
\E[\bone_{\hat{Y} \neq Y}] &=& P(\hat{Y} \neq0) (1-p_y) \\
&&{}+ P(\hat{Y}
\neq1)p_y =q_y(1-p_y)\\
&&{} +
(1-q_y)p_y.
\end{eqnarray*}
So the best predictor will be the one that minimizes the above
expression. Taking derivatives with respect to $q_y$, we obtain
\[
1-2p_y,
\]
which is positive if $p_y < 1/2$ and negative if $p_y > 1/2$. So, if
$p_y <1/2$, the minimum error is achieved by taking $q_y = 0$ and if
$p_y >1/2$, the minimum error is achieved by taking $q_y =1$. This
intuitively makes sense; if the loss is 0--1, then make the predictor
the outcome with the highest probability. So if we take $\hat{Y} =
\argmin_y\{P(Y = y)\}$, then that predictor has the lowest 0--1 loss
and its error is given by
\[
\E[\bone_{\hat{Y} \neq Y}] = \min\bigl\{P(Y=0), P(Y=1)\bigr\}.
\]
The same argument implies that best predictor when conditioning on $X$
is $\hat{Y}(X) = \argmin_y\{P(Y = y|X)\}$ and the error is
\[
\E[\bone_{\hat{Y}(X) \neq Y}|X]= \min\bigl\{P(Y=0 |X), P(Y=1 |X)\bigr\}.
\]

To see that $I$ has the desired properties, simply define a function
$g(0) = 0$ and $g(x) = 1$ for any $x \neq0$. Then $g$ is a
nonnegative penalty function and since $\bone_{\hat{Y} \neq Y} = g(Y
- \hat{Y})$ we can use the same machinery from the prediction section,
and our information link function therefore satisfies the assumptions
of Definition~\ref{linkfunc}.


\subsection*{Magnetic Substorm Dependence Estimation}
We consider the problem of estimating the functional dependence measure
given data $Y_1, \ldots, Y_n$ and $X_1,\ldots, X_n$. We assume that
$X_i$ and $Y_i$ are functions taking values in $L^2[0,1]$, are centered
and satisfy a functional linear model, that is,
\[
Y_i(t) = \int\beta(s,t) X_i(s) \,ds+
\varepsilon_i (t),
\]
and that $\beta$, as an operator, is bounded. We assume that both $\{
X_i\}$ and $\{\varepsilon_i\}$ are i.i.d., are independent of each other, and
\[
E\|X_i\|^2 < \infty\quad\mbox{and}\quad E\|\varepsilon_i
\|^2 < \infty.
\]
For the estimation of $D_1$ and $D_2$, we refer to (\cite{rasi2005}).
The third measure, $D_3$, we have not seen in previous literature so we
provide a consistent estimator. Define $C_Y(s,t) = E[Y(s)Y(t)]$ and
$C_X(s,t) = E[X(s)X(t)]$. Assume that the first $d+1$ and $q+1$
eigenvalues (ordered by magnitude) of $C_Y$ and $C_X$, respectively,
are distinct. Define the projections $\bY_1, \ldots, \bY_n$ and $\bX_1,
\ldots, \bX_n$, where
\[
\bY_{i,j} = \langle Y_i, \hat{u}_j \rangle
\quad\mbox{and}\quad \bX_{i,k} = \langle X_i, \hat{v}_k
\rangle
\]
for $j=1,\ldots, d$ and $k=1,\ldots, p$. Here $u_j$ is the $j$th
eigenfunction of $C_Y(s,t)$ and $v_k$ is the $k$th eigenfunction of
$C_X$ ($\hat{u}_j$ and $\hat{v}_k$ are the sample counterparts).
Notice that $\bY_{i,j}$ is uncorrelated across $j$ since we are
projecting onto the eigenfunctions of $C_Y$. Therefore, the $D_3$
measure can be expressed as
\[
D_3(X,Y) = 1 - d^{-1} \sum_{j=1}^d
\frac{ E(\langle Y, u_j \rangle-
E[\langle Y, u_j \rangle| X])^2 }{ \lambda_j^2},
\]
where $\lambda_j$ is the $j$th eigenvalue of $C_Y$. We can express
$\bY_{ij}$ as
\[
\bY_{i,j} = \bbeta\bX_i + \bdelt_{i,j} +
\beps_{i,j},
\]
where $\bbeta_{j,k} = \langle\beta, \hat{u}_j \otimes\hat{v}_k
\rangle$, $ \beps_{i,j} = \langle\varepsilon_i, \hat{u}_j \rangle
$, and $\bdelt_{i,j} = \sum_{k= p+1}^\infty\bbeta_{j,k} \langle
X_i, \hat{v}_k \rangle$. If we define
\[
\hat{\bbeta} = \bigl(\bX^T \bX\bigr)^{-1} \bX^T
\bY,
\]
then one can show that $\bbeta- \hat{\bbeta} = o_P(1)$, $\hat{v}_k -
v_k = o_P(1)$, and $\hat{u}_j - u_j = o_P(1)$; see \citet{hokore2009}
for details. Define the fitted values $\hat{\bY}_i = \hat{\bbeta}
\bX_i$. We then estimate $D_3$ as
\[
\hat{D}_3(X,Y) = 1 - d^{-1} \sum
_{j=1}^d \frac{1}{n}\sum
_{i=1}^n \frac{ (\bY_{i,j} - \hat{\bY}_{i,j})^2 }{ \hat{\lambda}_j^2}.
\]
It is then easy to show that (via Slutsky's lemma and the law of large
numbers)
\begin{eqnarray*}
\hspace*{-4pt}&&
\hat{D}_3(X,Y) \\[-2pt]
\hspace*{-4pt}&&\quad\stackrel{P} {\to} 1 - d^{-1} \sum
_{j=1}^d \Biggl( E\Biggl[
\langle\varepsilon_{1,j}, u_j \rangle\\[-2pt]
\hspace*{-4pt}&&\hspace*{98pt}{}+ \sum_{k= p+1}^\infty\langle
\beta, u_j \otimes v_k \rangle\\[-2pt]
\hspace*{-4pt}&&\hspace*{148pt}{}\cdot\langle X_1, v_k \rangle\Biggr]^2 \Big/
\lambda_j^2\Biggr).
\end{eqnarray*}
Notice that ideally we do not want the term
\[
\sum_{k= p+1}^\infty
\langle\beta, u_j \otimes v_k \rangle\langle X_1, v_k \rangle
\]
in the above expression, but that is the error we make in projecting to
a finite dimension. But, if we choose $p$ large, we can make the term
arbitrarily small and, in practice, we expect that term to contribute
relatively little to the overall estimate.
\end{appendix}

\section*{Acknowledgments}

The authors would like to thank Xiao-Li Meng, Radu Craiu, Stefano
Castruccio, Ryan King, three anonymous referees and the Associate
Editor for comments that have greatly improved the manuscript.


%

\end{document}